# An Empirical Investigation of Soft Skills and Corresponding Challenges in International Software Engineering Education

Sajid Ibrahim Hashmi[0009-0004-8866-1280], Jouni Markkula[0000-0003-1075-5303]

M3S Research Group, Software Engineering and Information Systems Research Unit, University of Oulu, Finland
{sajid.hashmi,jouni.markkula}@oulu.fi

**Abstract.** Joint software development work requires the application of technical skills and professional practices. Soft skills, such as teamwork, remain essential to succeeding in a professional software development environment. The phenomenon is reflected by group work in SE (software engineering) education, where students can learn and practice teamwork-related soft skills. Therefore, SE education requires students to practice those skills so that they can actively embark on their professional careers in the software industry.

In this research study, empirical findings from two focus groups we conducted on group work are presented, with emphasis on the soft skills necessary for teamwork in international software engineering education, specifically student collaboration in higher education. There are several soft skills students should learn as they practice teamwork in their course-related assignments and exercises, as well as the challenges that come with it in international settings. The findings serve as a guideline to university teachers and pedagogy experts to plan student group work more effectively.

## 1 Introduction

In general, teamwork helps achieve tasks beyond the capacity of individuals, and teams tend to perform better for the same reason [1]. The joint work requires the application of technical skills and professional practices. Teamwork is an essential practice for succeeding in a professional software development environment. The SE team members must possess the necessary skills and required competencies.

SE is a teamwork activity in which software engineers collaborate as software is developed in a professional environment. Software engineers should be able to collaborate and interact effectively and proficiently with other software engineers, customers, and users to carry out development activities. SE is also an international discipline, as software companies and other organizations are often international and have multicultural personnel. Therefore, working in teams is not trivial, as it requires specific professional capabilities and skills from participating team members.

Teamwork skills are highly relevant in industry environments [1], and students' participation in teams for group assignments can help them gain relevant experience [2].

Not only that, group work in SE education prepares students to collaborate with future colleagues, which becomes easier and more efficient when they can apply and consolidate the knowledge gained at the university [3]. That, in turn, prepares work-ready graduates by systematically requiring them to work on SE-related coursework.

The study investigates teamwork-related soft skills in international SE education and the relevant skills students can learn and practice through group work in course-related assignments. The findings explore the topic through the teachers' perspective, as they are responsible for guiding students. Therefore, the study's outcome should guide SE teachers in designing SE coursework not only to have students practice teamwork-related soft skills but also to help them better understand the phenomenon's underlying motivation.

To support the study, focus group methodology was chosen to obtain relevant information from experts, including SE university teachers. The international SE education and related aspects were set as the study context. The overall purpose is to study the skills students can learn through group work and better understand how teamwork training should be made an essential part of international SE coursework.

The rest of the paper is organized as follows. Section 2 reviews related research on international SE education and teamwork; Section 3 details the research methodology; Section 4 presents the empirical findings from the focus groups; and finally, Sections 5 and 6 cover the discussion and conclusions, respectively.

## 2 Related Work

This section reviews the literature relevant to the study, providing theoretical and empirical background. Since the study fetches multiple concepts together, our review of the literature is multifaceted. It first reviews soft skills through some existing taxonomies; it then examines the growing importance of soft skills in the wake of the internationalization of higher education. Finally, it reflects on teamwork skills within international software engineering education, highlighting their significance.

### 2.1 Soft Skills and Existing Taxonomies

Soft skills are essential for young software developers to secure a position and begin their careers [4]. As with definitions of soft skills, their taxonomies also vary, and there is no consistent definition in the existing literature [5]. For instance, Garousi et al. [6] treat 'communication' as distinct from teamwork, although in general the practice is considered an essential part of successful teamwork. Furthermore, [6] considers soft skills to comprise team and interpersonal skills; Cinque [7], on the other hand, holds that soft skills comprise personal, social, and methodological skills. [8] has also categorized communication as a soft skill, independent of teamwork, which can complement the former. In practice, the existing literature offers no constraints on the use of non-teamwork soft skills in teamwork, suggesting that individual skills can also affect how a team performs as a whole.

A trait of the existing literature on competency and skills is that the terms competency and skills are used interchangeably, with no clear distinction drawn between the two concepts. For instance, Niva et al. [4][11] and Assyne et al. [12] has grouped phenomenon such as teamwork, problem solving, communication, negotiation, and resilience, to name a few, as competencies formed by a combination of soft skills, and, on the other hand, the software engineering competency model [9], and [5] [13][6][14] have listed the same phenomenon individually as skills. Therefore, in our work, we do not aim to create a disparity between "soft skills" and "competencies" and use both terms interchangeably.

### 2.2 Internationalization and Need for Soft Skills

With advances in communication methods and technologies, terms such as internationalization and globalization have also become common in the education sector. Although the concepts are linked, they can be interpreted in multiple ways depending on the context. In terms of higher education, the terms' globalization' and 'internationalization' can be defined as *"the flow of technology, economy, knowledge, people, values, ideas, across borders"* and *"the ways a country responds to the impact of globalization yet, at the same time respects the individuality of the nation"* [15] respectively. Although viewed differently, these are linked concepts; globalization can be considered the catalyst, while internationalization is the proactive response [15].

International education has become a central focus of educational research, with globalization and the internationalization of higher education as distinct areas of inquiry. Each approach is linked to the global, social, political, economic, and cultural shifts that shape the overall approach [16]. It has also been reported in the literature that institutions need to adopt specific approaches, in addition to any existing generic approaches in practice, to respond to the internationalization of higher education, such as activity, competency, ethos, and process [17].

Internationalization or globalization of education is one-way; professional students, who make up a reasonably significant segment of a community, can be equipped with better skills to compete with these phenomena and the rest of the world. That is not to suggest that hard skills are irrelevant or no longer important; rather, the global economy and the functioning of international organizations require multi-skilled graduates. The requirement is becoming increasingly relevant in the current international landscape. It is worth clarifying that in this study, the phrase "international SE education" refers to teaching international students in a particular local context.

### 2.3 Teamwork Skills in International SE Education

Software development teams have become a workforce with diverse cultural and educational backgrounds due to their social nature, which entails the involvement of multiple team members [18][19]. Likewise, in international SE education, students come from diverse backgrounds. In any case, it is not possible to achieve the benefits of collaborative work without effective teamwork within teams.

From an international education perspective, millions of STEM graduates are produced worldwide, and many lack adequate soft skills to meet global market demands [20]. Soft skills, such as teamwork, are essential to meet the pressing needs of the international work environment [21]. However, in some countries, soft skills do not make a substantial part of secondary and tertiary education [22].

Teachers and students are both responsible for learning, and through making students work in groups, following advantages can be achieved [23]: Group work may engage students in what they are learning for internalization to occur through repeating or sharing the contents; the phenomenon allows students to distribute different types of tasks among themselves; individuals can be more motivated when offered an element of choice in terms of the type of work they want to do. Cultural background and beliefs are likely to impact learners' behaviour, interpretation, and understanding [23]. It is also the case that conceptions of learning may differ between two countries, reflecting cultural differences [24].

However, it has been reported that software companies have noted that graduates, despite their technical expertise, lack the skills needed to work in groups [6], underscoring that soft skills are as crucial as technical capabilities for students to succeed in their professional careers. Group work, on the other hand, is a recommended way for students to practice the professional skills required in the industry [25]. However, introducing group work alone in SE courses is not sufficient unless it is implemented effectively. The existing work on SE education mainly addresses content delivery in a globally distributed setting and the overall challenges that come with it [26][27][28]. Yet, it fails to address the issues that confront students' collaborative learning in international settings.

## 3 Research Methodology

Considering the research objectives and motivation, the research questions were formulated as follows.

**RQ1:** What teamwork-related skills do students learn through group work in international software engineering education?
**RQ2:** What teamwork-related skills can students learn through group work in international SE education?
**RQ3:** What are teamwork-related challenges in international SE education with respect to culture?

The first two questions might look similar, but their interpretations differ. RQ1 is based on teachers’ experiences, such as the skills they are currently teaching and what they think students are learning from their existing teaching, as observed in student group work, which prompts them to assign group-based tasks. On the other hand, RQ2 aims to investigate which skills students currently lack, which they can learn with teaching support, and which they should be learning in addition to the existing ones. The question was also aimed at professional skills, that is, the soft skills required by the software industry for them to practice after graduating at a different level. That said, some of the skills acquired from RQ1 and RQ2 might overlap; however, for RQ2, participants had more context and did not have to rely on memory because they were provided with a sample list of soft skills drawn from multiple scholarly sources. For the same reason, their selection of the soft skills students should be taught through group work was more explicit and included comparisons and examples that linked to their own teaching. In RQ3, we address cultural challenges in the subject under discussion.

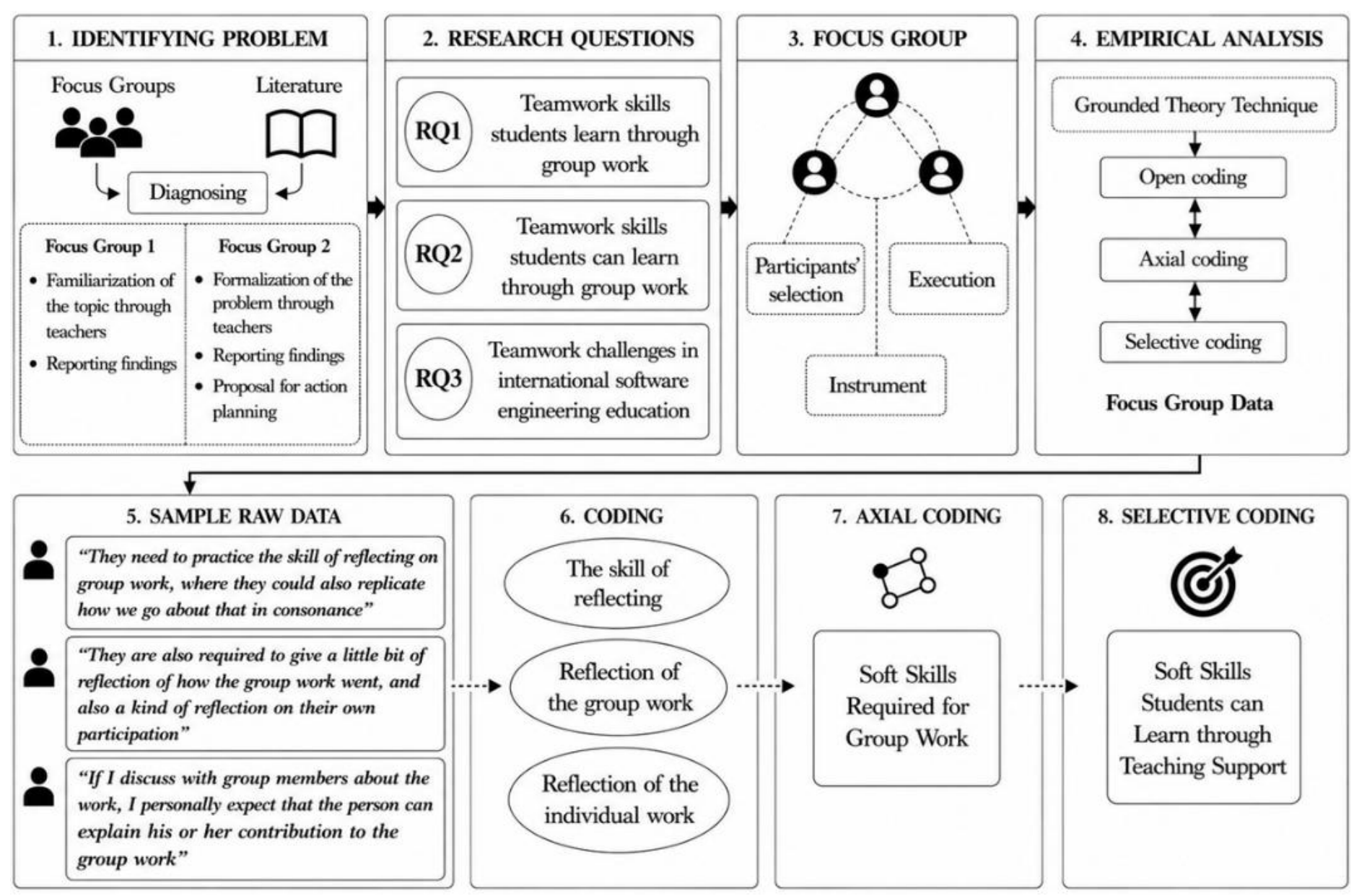

**Fig. 1.** The Overall Research Process

Action research [33] was chosen as the research method for the overall research project. However, this study presents the first phase of the Action research cycle, called Diagnosis, during which we conducted two focus groups to investigate the problem, a well-established method for collecting empirical data [29]. The university teachers, with experience profiles in Table 1, were chosen as subjects because they possessed strong knowledge and understanding of various teaching methods. The empirical data and results presented in this article are derived from two sub studies. The first was conducted as a pilot in the fall of 2022, and the second was conducted in 2025, a continuation of the first to make the overall study longitudinal. The participants gained additional international experience in between through an SE education exchange program with a university in China. The pilot focus group, FG1, was attended by three SE teachers, and the second focus group, FG2, was attended by ten teachers at the University of Oulu. All participants taught international and local groups within the country and abroad. The execution of the pilot focus group helped tweak the questions asked of

**Table 1.** Experience Profile of the Study Participants

| SE experience (years) | Overall experience (industry + teaching) | International settings | |
|---|---|---|---|
| | | Overall | Teaching |
| 1–5 | 1 | 2 | 2 |
| 6–10 | 3 | 3 | 4 |
| 11–15 | 5 | 4 | 3 |
| 16–20 | 1 | 1 | 0 |
| ≥20 | 2 | 2 | 1 |

**Note.** Values indicate the number of participants.

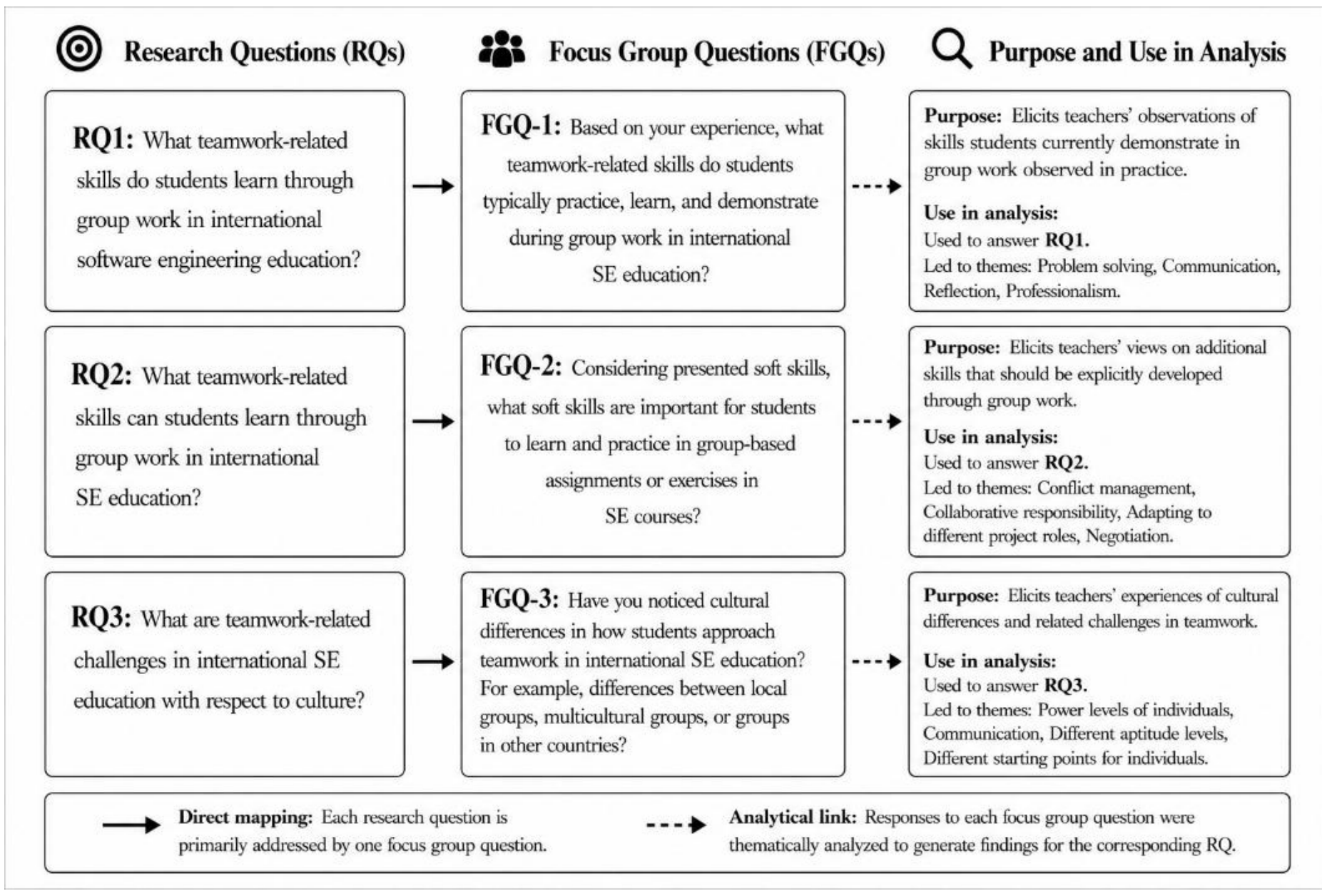

**Fig. 2.** Link between research questions and focus group questions

the experts, with a view to acquiring more targeted data.

In our endeavours to answer the research questions, participants were engaged through open-ended questions, presented in Figure 2, along with how each linked to the corresponding research question, their purpose, and their use in the analysis. To support participants' memory, they were shown three candidate lists of soft skills to help them answer the focus group questions furnished in the Appendix. For the pilot focus group, only list C was made available, as the authors had limited context for the problem domain and also wanted to learn how to ensure the focus group ran smoothly. However, participants were encouraged to bring soft skills not on the list. For analyzing qualitative data, thematic analysis was chosen, a foundational method for qualitative data analysis [30]. The first author analyzed and coded the data, and the codes were discussed with the second author in multiple meetings to reach an agreement.

## 4 Results

This Section presents the research results. For the sake of clarity and readability, we follow the same investigative scheme for presenting the results, organized by our research questions: the teamwork skills students currently learn, the ones they should be provided support for, and the challenges that come with practicing the phenomenon, with an emphasis on culture in the international education setting. Figure 3 depicts the overall framework as a branching model, based on empirical findings from our study, that links the identified skills and challenges to the research questions. It is worth noting that, due to space limitations, we are only partially reporting the results of the study we conducted. In terms of skills, the crux of the findings is that the teachers distinguished between skills that students practice naturally through group work and those that require instructional support.

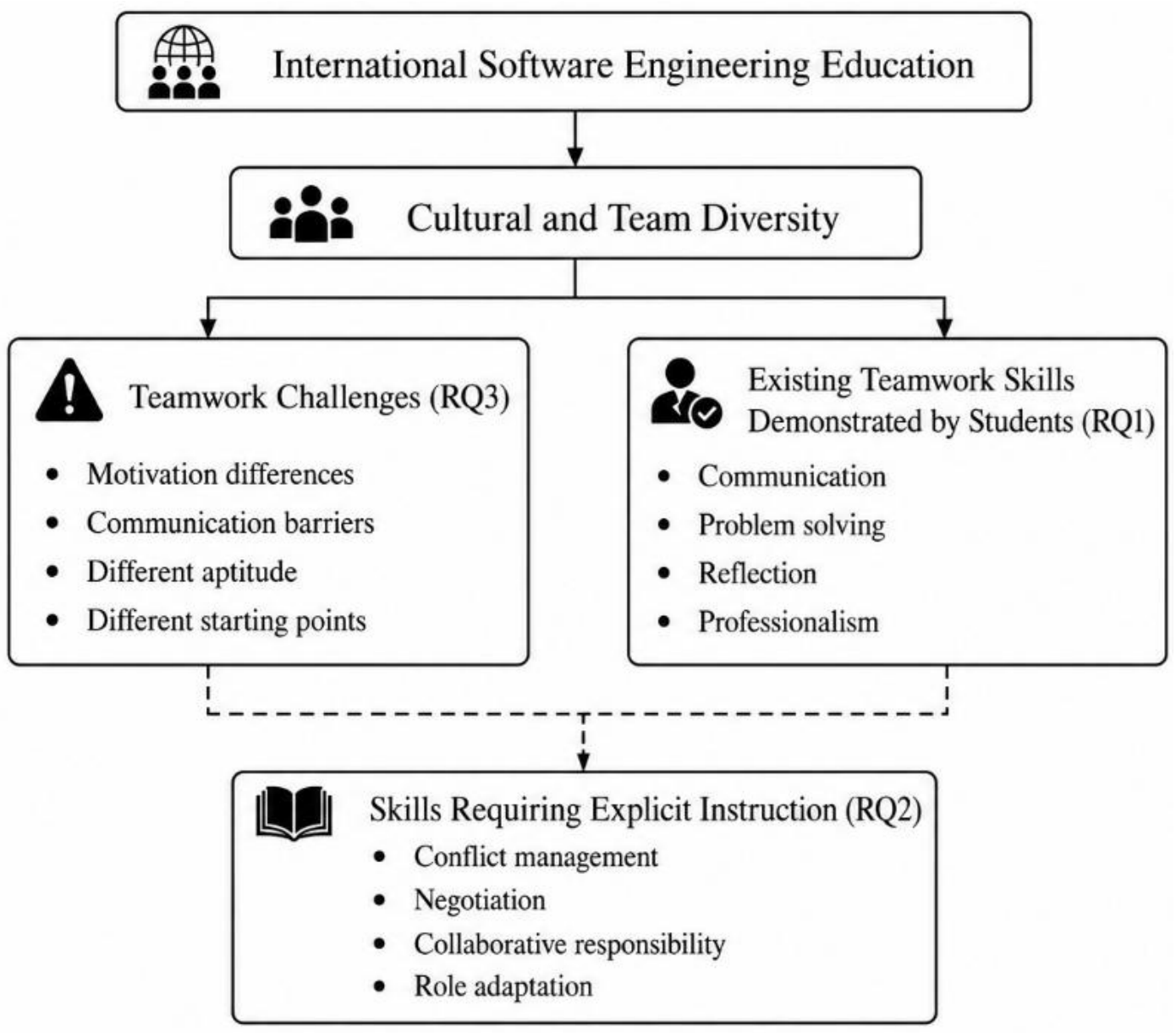

**Fig. 3.** The Overall Framework linking challenges and skills

### 4.1 Teamwork-related skills students learn through group work (RQ1)

Analysis of the focus groups that the study participants consistently observed the following competencies listed in the section as the ones that naturally emerge through group work. These are outlined as outcomes of collaborative work rather than skills requiring instructional intervention. In a nutshell, the skill themes together represent a category of skills that emerge naturally.

**Problem solving**. Problem-solving entails selecting the correct solution from several options through reasoning and decision-making [31]. One advantage of group work is that it helps students learn problem-solving, a core requirement of the software industry, and replicate it while they are studying. In the Focus group, the experts reported that learning to solve problems was an important lesson in teamwork. If done together, it helps solve a problem more effectively. An expert, who had an industry background as well, expressed,

> *"There are different ways of learning, and they do individual learning as well, but real problem solutions, let's say that, typically programming work in industry is done by a person, but problem solving is done together."* (FG1)

While working in a team, students also do individual learning. However, problem solving, which is core to working in the software industry, is always done collaboratively, and teamwork on course-related projects provides students with an opportunity to learn the skill. Another participant had a similar view of the skill for student teamwork.

*"I try to emphasize that it is often easier for them to start solving the problems if they do it together."* (FG1)

It is evident that one participant viewed practicing problem-solving as a software industry requirement and a way to prepare students for their professional careers. In contrast, the other participant thought that practicing problem-solving collectively facilitates students' problem-solving by allowing them to share their knowledge. The perspectives might differ, but the motivation to practice and learn the skill remains similar.

**Communication.** It is widely considered a soft skill, whether or not it is part of teamwork. Teamwork in SE education provides students with opportunities to practice communication, collaboration, negotiation, and conflict resolution, thereby developing these skills for work-related situations in industry. The IEEE's Software Engineering Body of Knowledge (SWEBOK) [33] knowledge area, Software Engineering Professional Practice, is divided into three subareas, one of which is communication skills: "Professionalism"; "Group Dynamics and Psychology"; "Communication Skills". Teamwork is explicitly presented under "Group Dynamics and Psychology", as "Dynamics of Working in Teams/Groups". "Communication Skills" is also closely connected to teamwork-related skills. In line with the SWEBOK's taxonomy, one of the experts asserted that communication should be professional. Not only that, communication is linked to making students learn to present a project plan and results, e.g.,

*"But there is also this aspect of professionalism, so they have to practice the professional communication; towards the customer, towards the teacher, and a kind of learning to present the project, the plan, and the results."* (FG2)

Communication-related skills, such as presentation, came up as important expertise that team members can develop, as was revealed from the discussion with experts, e.g.,

*"You need to make it clear in the evaluation that everyone needs to be presenting, and it depends on them how they want to do the presentation. Nobody is good at it at first; we have to practice it to do it."* (FG1)

Presentations are one way of professional communication. Moreover, the phenomenon involves students developing writing and documentation skills before they can stand up and deliver. That, in turn, requires students to be good listeners and understand each other well in a group before they can present themselves.

**Reflection.** Reflection as a group helps team members evaluate others' work, thereby improving collaboration and fostering collective problem-solving [34]. It plays an important role in promoting learning and is reported to have multiple benefits: it is an account of how students confronted challenging situations, draw on experience to improve for the future, brings different ideas and enhances learning, helps learners see the interconnections of the knowledge learning, and makes students find social connections among their fellows [35]. In our study, reflection came up as a recurring topic as a means for individual and inter-team evaluation, e.g.,

*"Analyze their own work, what is good and what did not work. Evaluate others' work as well, and inter-team evaluation, their reflection on their own work."* (FG1)

Students must learn to reflect on the group's overall work and their own participation. Doing that can help them replicate their success. One participant said they use the learning diaries for this purpose, which helps teachers better understand the scenario for multicultural student teams. Studies, for example, [35], suggest that instructors can design prompt/relevant questions for the reflection to improve learners' meta-cognitive knowledge, e.g.,

*"We provide some questions for them to reflect on, which also helps us to evaluate."* (FG2)

Reflective practices are known for enhancing teamwork and individual learning [36]. Therefore, teachers' roles can help them develop soft skills relatively quickly. For instance, upfront notice that students will be evaluated independently and in the middle of their work can help them understand what they are expected to do, which, in turn, can improve their planning skills.

**Professionalism.** Teamwork involves implementing professional practices alongside technical skills. Certain characteristics of the team and its activities may affect teamwork operations and productivity. Success depends on methods and tools, and it also heavily relies on how software engineers are educated to incorporate professional practices [10][18]. This practice repeatedly came up, and one of the instances of professional growth was that they must be able to behave responsibly not only within their group but also to the other groups as well, and it starts with communication, as one of the participants said,

"*Speaking of professionalism, also not towards the customers and teachers, but also like professionalism while working as a group...and towards the other group is also important."* (FG2)

It was agreed among the participants that students must be able to practice professionalism when working as a group and when working with other groups. Moreover, professionalism fosters professional growth among group members, and the meaning of that growth may differ. They should learn to present the project in terms of plans and results, as those are evaluated against the given set of requirements as part of group work.

**Key insights from Section 4.1:** Problem-solving is an essential skill for a professional career, and teamwork provides students with an opportunity to practice it. Communication is not about speaking alone; it remains insufficient unless it is professional and is exercised towards all stakeholders. The skill is closely linked to plan generation and to presenting project results, which are crucial to an individual's ability to reflect on themselves.

### 4.2 Teamwork Skills Students Should Learn Through Group Work (RQ2)

Unlike the skills identified in RQ1, the study participants identified the following skills as requiring pedagogical support. The participants agreed that simply assigning students to multicultural or random teams was not enough to help them develop those skills.

**Conflict Management.** An important aspect of conflict management is that if any group member is not contributing as expected by the group or in line with the group's work demands, they should not be excluded; instead, they should be engaged, and students should not come complaining to the teacher at the end. They need to discuss among themselves. One of the experts expressed that,

> *"There are also these groups where there might be some conflicts inside the groups, so that some person does not work, for example, so this conflict management is also something that I would like somehow that they need to learn; not that after the group work you hear that these two people did not do it, so we excluded them. That is not like you were supposed to do; you were supposed to discuss it among yourselves somehow."* (FG2)

Conflict management is challenging in multicultural settings for two reasons: the language barrier and students' inability to address their grievances promptly, both of which may be influenced by cultural factors. For example, in some cultures, harmony is expected, while in others, people fight to resolve conflicts. For these reasons, the timely resolution of work-related conflicts is crucial to the successful execution of group work.

**Collaborative Responsibility.** It has to do with responsibility too, as the group project should make them responsible, which is a collaborative responsibility that falls under professional ethics and is how they should be performing, individually as well as in a group. As it came up during the focus group,

> *"I think that is the responsibility part here that once they are in a group project, they should be responsible for their own part, and I think it is also to this 'collaboration' that they are a kind of collaboratively responsible for their group work, if so, in that sense they have been under this 'professional ethics'. However, then this is how they should be performing at the individual level, but also together as a group."* (FG2)

Along with collaborative responsibility, responsibility alone emerged as a general theme that should be present at multiple levels to support student teamwork. On the other hand, the perception of a lack of responsibility on the part of teachers could demotivate students from embracing the core philosophy behind group work. As one of the participants said about the perception of the students,

> *"They would undermine their teacher's commitment to accountability. They deserve the feedback for what they have done, so they must all be evaluated."* (FG1)

Upon inquiry, the same participants elaborated that it could be the overall academic culture, or the culture of favoring the good students, that is even undermining the not-so-good ones. Essentially, it is not only students who should behave responsibly; teachers, in their role, should as well, so they can gauge teamwork effectively.

**Adapting to Different Project Roles.** Courses or modules focused on software development require students to take on different project development roles, providing an opportunity to practice the soft skills associated with each role. Teachers can help the students choose suitable roles for themselves, especially those who do not have much experience with teamwork, e.g.,

*"We have this Software Organization module, for example, we take an agile team, then they have these different roles working together in a group to develop a product, actually. What we do is like we try to help the students to take these different roles doing the project work because each role has different soft skills to maintain that work and then synchronize that thing."* (FG2)

For instance, if the course is about agile software development and students are being taught Scrum, they are expected to learn multiple roles so they can practice a range of soft skills, as each role requires specific soft skills to execute effectively. The participant also mentioned that, in a few instances during the course, they allowed students to take on self-organized team roles.

**Negotiation.** It requires individuals to be flexible enough to change their minds and willing to negotiate. A participant emphasized the importance of negotiation skills and said,

*"So these negotiation skills could be very much highlighted that you need to negotiate your goals and bring your own goals to the table and agree with the group."* (FG2)

It implies that effective practice of negotiation skills is crucial for achieving goals, as the preliminary stage is identifying them, and the skill is needed to set them up. It is not just about setting individual or group goals; negotiation skills are also required to carry out tangible project-related tasks, especially when students are given the freedom to choose their topics. Negotiation may help them reach an agreement in cases of conflicting choices, as one expert from the same focus group shared their experience:

*"I think some of the skills they would be needing are negotiating skills because sometimes it gives them the freedom to choose the topic they will be working on if it is a group assignment or something."* (FG2)

It implies that flexibility and negotiation are linked to the extent that, to negotiate well, one should be flexible enough to listen to and accommodate others.

**Key insights from Section 4.2:** Students from diverse backgrounds may handle and negotiate conflicts differently. The phenomenon makes them learn how to deal with conflicts with their culturally diverse teammates. The need or urge to address conflicts should make them collaboratively responsible for their project task. The 'onus of responsibility' should be on teachers as well to treat each team member equally and to provide timely, equal feedback on their work. Each role in software development projects has certain soft skills associated with it. Switching student roles, for instance, Scrum-based development courses can help students not only adapt to different project roles but also learn and practice skills associated with each role.

### 4.3 Challenges in Terms of Effects of Culture on Skills (RQ3)

The challenges identified in the research question explain why study participants, as teachers, considered instructional support necessary for students to learn those skills. Many difficulties in teamwork were attributed to individual differences in motivation, aptitude, and starting points, as well as to differing communication styles.

**Varying Levels of Motivation.** A varying level of motivation in a multicultural group is one thing that different cultures can inspire. If motivation is considered a skill, it appears that there are some factors that, in a way, might be culturally dependent and can make students more motivated, regardless of their cultural background; for instance, students travelling to a different country and studying in its environment, which requires acquiring and disposing of means. As the participant was quoted as saying that,

*"One of the things we have noticed is that students have different motivations regarding the course; someone just wants to pass the course, and then someone wants to get high grades. That motivation actually drives them to be more efficient or less efficient. However, if we have this kind of combination of students, this actually creates problems or even challenges to conduct this teamwork."* (FG2)

That said, motivation is more of an individual characteristic and is not culture-dependent in its true sense. The same is suggested by motivation theories, such as the content and process theories, which list multiple sources of the phenomenon; culture is not among them. Therefore, it is worth clarifying that the motivation here is only to get good grades, which can be considered extrinsic motivation. Some students might want to get grades by doing nothing, and others' objective would be different.

**Communication.** Communication is a basic skill essential for teamwork, and students may practice it as much as any other skill. However, the phenomenon also poses challenges in multiple ways: reduced expressiveness, language barriers, differences in communication styles, and the need for a common language for effective communication. Communication style may vary from formal to informal, and can adjust team members' expectations for how they are treated. One of the experts expressed their opinion that,

*"In some cultures, their communication style is inherently different, and it shows in communication in every way they formulate the emails that they send."* (FG2)

The language barrier came up as one of the things that prevents students not only from communicating with each other but also from asking the teacher questions. Another expert shared their experience teaching in the international environment, e.g.,

*"Not only that, but you know, even when you allow students to open up and say that if you are having any challenge, please communicate it now. They will not be telling you that during the sessions, when everyone is there.."* (FG2)

In that regard, the teacher's role is equally important in giving them the confidence and opportunity to be open and express themselves. At the same time, not only is it the common language, but it is also a communication challenge at a lower level within the same domain, as people get used to different dialects and/or accents.

**Different Aptitude Levels.** In addition to varying motivation levels, individuals in a student team may have different aptitudes, which can affect teamwork both positively and negatively. Positively, because those individuals would strive to do their best in their roles, which in turn would improve overall teamwork. On the contrary, having one or two individuals with higher or even lower aptitude levels than the rest may make the rest of the team members very challenging to catch or even cope with the former, e.g.,

*"If a person comes from a very competitive aptitude, and sometimes some cultural aptitude could be competitive, for example, sometimes your aptitude is that I want to be the best. So that kind of thing can make a difference in this thing."* (FG2)

However, it is worth noting that competitiveness is not uniform across countries. Culture may influence how individuals compete, as traits can be rooted in different social layers. For instance, a study by Hauge [37] asserts that cultural norms transmitted through families, particularly by parents who had moved to different countries, can positively influence competitive behaviour. Different aptitude levels among peers in a group can lead to the prevention of knowledge sharing. For instance, students with high aptitude might want to do everything on their own, without depending on their peers' contributions.

**Different Starting Points for Individuals.** Another interesting point that came up for international SE students is different starting points considering different backgrounds of the students, e.g.,

*"So I think there are those, but what a culture is, is a very difficult thing to say and to point. But if there are very competitive starting points for individuals, that is something."* (FG2)

It implies that teachers would need to start differently for individuals from different international academic backgrounds. Those starting points could be influenced by language barriers or differences in the individuals' aptitude or motivation levels. Teaching arrangements can also pose different situations; for instance, local students and international teachers, and vice versa. As one of the experts said,

*"But what is interesting in China is that there the students are from the same country, and we are just the international teachers. So that is a slightly different situation."* (FG2)

The scenario poses a significant challenge as well. For instance, teachers from another academic culture might find it challenging to adapt to students' culture and even feel alienated from it. In that situation, the roles of teachers and even institutional management may also come into play, depending on whether they prefer students to be taught in accordance with the existing academic culture or encouraged to adopt a new one.

**Key insights from Section 4.3:** Just as benefits can be composite, challenges can be as well. There may be varying levels of motivation among students in a team with diverse members, driven by factors such as the desire to earn grades and the decision to relocate to another country to acquire knowledge, which may also influence their aptitude levels. Furthermore, communication styles or stereotypes among such diverse members would differ. All in all, that makes teamwork very challenging. Teachers would also find it challenging to guide a diverse group of students with different starting points, varying aptitudes, and/or motivation levels.

## 5 Discussion

In light of the existing research, we conducted focus groups to investigate the domain further; the results are presented in Section 4. The Focus groups with experts provided significant insights into teamwork skills in international SE education. Students can learn multiple teamwork skills through group work in their course-related assignments. The findings are summarized below in light of the research questions.

*RQ1-What teamwork-related skills do students learn through group work in international software engineering education?*

The existing work on teamwork skills or challenges has been course-specific, or has been applicable in a given settings or with a limited context, for instance [38][39], or mapping or review studies [40][41] which mainly rely on the frequency of the occurrence of various soft skills in the scholarly articles to manifest the degree of importance of each skills, instead reflecting its true value in today's academic settings by incorporating the actual stakeholders, the teachers, and their real world relevant teaching experience. In international settings, the usefulness of some soft skills, their context, and the roles they play can be even more important, and they can help students further reap the benefits of collaborative work. The international environment from which students come, with its diverse social and academic cultures, provides them with an opportunity to develop their skills. Although teamwork may involve individual work, it still requires problem-solving, which is how things are done in the software industry. The ability to communicate well is a known soft skill. However, professional communication is more beneficial for students' learning, as it is required for presentations and helps them reflect effectively on themselves when needed. That eventually leads to adopting professionalism in general, which serves as the source of practicing multiple soft skills such as collaboration, negotiation, and conflict handling. A combination of skills, such as problem-solving, reflection, and group learning, can augment and lead to the learning and practice of other skills as well; for instance, cooperation and coordination require communication, and vice versa. Knowledge sharing may foster cooperation and work planning, which may help develop plan generation skills. Teamwork may also help the members learn to deal with peers from different academic backgrounds. It is worth mentioning that this study was not aimed at producing a unique set of soft skills or a new taxonomy of the phenomenon. Our findings are consistent with the existing literature, at least regarding naming conventions.

*RQ2-What teamwork-related skills can students learn through group work in international SE education?*

There are multiple teamwork-related soft skills that students should learn through group work, especially as they prepare to enter the software industry and compete for professional careers. Those include, to name a few, conflict management, shared responsibility, adaptation to different project roles, and negotiation with one another. Such skills are closely linked to each other. It is the recurring challenges associated with culturally diverse student teams that drive the need for these skills. On the other hand, existing research on project-based software engineering courses that require group work overlooks the importance of many soft skills and focuses primarily on the more commonly recognized ones. Not only have we listed the soft skills, but we have also argued how they complement each other. For instance, effective conflict management does not advocate excluding conflicting voices and putting the onus on them, but rather engaging them. The practice should make students collaboratively responsible, at both the individual and group levels, for advancing the team's overall agenda. Teamwork in the project-based Scrum courses can be very beneficial for students to practice multiple skills: first, as a course requirement, it makes students well-versed at switching between different roles and adapting to them. Secondly, the practice may also help improve their communication skills by encouraging collaboration after they agree on a common language in international settings. Thirdly, changing project roles during the course of the project requires students to collaborate by designating their roles and to date work to their peers. The need to collaborate on those tasks makes them jointly responsible to one another and the team.

*RQ3-What are teamwork related challenges in international SE education with respect to culture?*

From a different perspective, the skills required of students in the international SE education domain can also be perceived as challenges if not catered for. While listing the challenges, we have also elaborated on the situations or scenarios in which teachers or students may encounter them. On the other hand, existing research in the domain has mainly focused on distributed [42] and global [43] aspects. It is limited to either a given context or the overall challenges, not the specific ones, for instance, teamwork. Learning teamwork-related skills in specific international settings can be significantly culture-dependent. At the same time, defining culture itself can be a tricky part because of the various factors that constitute or influence it.

For the same reason, cultural diversity among students in international settings can directly or indirectly influence multiple challenges, with varying levels of motivation and aptitude being highlighted as factors. Among other reasons, in international settings, an individual student's decision to travel to another country to study or acquire knowledge in a different culture can be a motivating factor that sets them apart from the rest. For the same reason, students' level of competency may also be linked to their cultural background, which can make some cultures appear more competent than others. Communication, which students need to practice extensively for successful teamwork, lies at the core of the problem due to the diverse attributes of individuals within a team, such as skill and experience levels, and this separation can be further widened by differences in motivation and aptitude within the same team. For similar reasons, individuals in a team may have different starting points, which poses a great challenge for teachers in guiding students and ensuring they have equal learning opportunities at the end of their teamwork.

## 6 Threats to Validity

Some factors can threaten the validity of the results; for instance, the study involved two focus groups of 13 teachers from the same university who teach international software engineering. A small number of participants in FG1, which was more of a pilot study, but evidence was drawn from its findings to support the conclusion. Moreover, varying levels of experience among participants may introduce bias into the results, with participants with more experience in teamwork-based teaching more likely to speak and participate in the discussion. However, speaking can be an individual personality trait independent of experience. Above all, the outcome of the focus group is regarded as the collective contribution of the group rather than that of any individual participant. Another possible threat to validity is that all participants are teachers from a single university, which limits generalizability to that context. Despite all efforts, the analysis, interpretation and synthesis of data may be influenced by researchers' personal biases. However, the themes identified from each focus group are consistent in terms of the nature of the skills and challenges, which lends credibility and significance to the reported findings.

## 7 Conclusion

This study aims to identify teamwork skills in software engineering (SE) education that are essential for students to learn and practice through group work in their course-related assignments and exercises. It is crucial to investigate teamwork, as professional SE education should prepare students to work in teams for their professional careers. To carry out the study, two focus groups were conducted with university teachers who have experience teaching SE courses in diverse international environments. The results were then theoretically analyzed not only to assess their usefulness for international SE education settings but also to strengthen them for use as input to the action planning phase of the action research cycle, from a broader research perspective, as part of the continuing future work.

The study revealed multiple teamwork skills students can learn and practice through group work in their SE education. Learning those skills can be helpful for students' future careers. The study also found that students from different cultures and academic backgrounds are likely to possess different work habits that not only pose challenges but may also help them learn diverse teamwork skills. However, cultural differences among teams can also affect the way students learn and practice those skills. The challenges in international SE education should guide teaching practices and prompt teachers to adapt their pedagogical methods and teaching content. All in all, the study findings can help SE teachers to design their courses accordingly and effectively execute teamwork.

**Acknowledgements.** The Grammarly tool was used to improve the language, and LLMs were used to polish the figures.

**Appendix.** Soft Skills Frameworks Presented to Focus Group Participants (Skills and/or competencies highlighted where relevant)

| |
|---|
| **Personal skills:** Learning skills, Tolerance to stress, Professional ethics, Self-awareness, Commitment, Life balance, Creativity/Innovation |
| **Social skills:** Communication, Teamwork, Contact network, Negotiation, Conflict Management, Leadership, Culture Adaptability |
| **Content-reliant/Methodological skills:** Customer/User orientation, Continuous improvement, Adaptability to change, Results orientation, Analytical skills, Decision making, Management skills, Research and info management |

**(a):** Cinque, M.: "Lost in translation". Soft skills development in European countries. *Tuning J. High. Educ.* 3(2), 389–427 (2016). [7]

| | Domain Name | Clusters of Competencies | Competencies |
|---|---|---|---|
| Domains of Competence | The Cognitive Domain | cognitive processes and strategies, knowledge, creativity | critical thinking, information literacy, reasoning and argumentation, and innovation |
| | The Intrapersonal Domain | intellectual openness, work ethic and conscientiousness, positive core self-evaluation | Flexibility, initiative, appreciation for diversity, and metacognition |
| | The Interpersonal Domain | teamwork and collaboration, leadership | communication, collaboration, responsibility, conflict resolution |

**(b):** National Research Council: *Education for Life and Work: Developing Transferable Knowledge and Skills in the 21st Century*. National Academies Press (2012) [44]

| Theme | Professional Practices/Skills |
|---|---|
| Communication (Coordination/Cooperation/Interdependence) | Documentation, speaking/explaining, listening, reading/summarizing/explaining, writing |
| Teaming | Consultation, decision making, discussion, plan generation, conflict handling, multitasking, multicultural knowledge transfer, commitment, accountability, group cohesion |
| Presentation | Lead reviews, product walkthroughs, training, documentation |

**(c):** Based on ACM Curriculum Guidelines [10] and IEEE-SWEBOK (Software Engineering Body of Knowledge) [33]